\documentclass[reprint,twocolumn]{revtex4}%
\usepackage{amsfonts}
\usepackage{amsmath}
\usepackage{amssymb}
\usepackage{charter}
\usepackage{graphicx}
\usepackage{float}
\usepackage{CJK}%
\usepackage{epstopdf}
\providecommand{\U}[1]{\protect \rule{.1in}{.1in}}
\begin{document}
\title{Quantum steering for two-mode states with Continuous-variable in laser channel}
\author{Kaimin Zheng$^{1,\ast}$, Jifeng Sun$^{2,\ast}$, Liyun Hu$^{2,\dagger}$, Lijian Zhang$^{1,\ddagger}$}

\affiliation{$^1$ National Laboratory of Solid State Microstructures, Key Laboratory of Intelligent Optical Sensing and Manipulation, College of Engineering and Applied Sciences, and Collaborative Innovation Center of Advanced Microstructures, Nanjing University, Nanjing 210023, China\\
$^2$ Center for Quantum Science and Technology, Jiangxi Normal
University, Nanchang 330022, China\\
$^\ast$These authors contributed equally to this work\\
$^\dagger$hlyun@jxnu.edu.cn\\
$^\ddagger$lijian.zhang@nju.edu.cn\\
}

\begin{abstract}
The Einstein-Podolsky-Rosen steering is an important resource for one-sided device independent quantum information processing. This steering property will be destroyed during the interaction between quantum system and environment for some practical applications. In this paper, we use the representation of characteristic function for probability to examine the quantum steering of two-mode states with
continuous-variable in laser channel, where both the gain factor and the loss
effect are considered. Firstly, we analyse the steering time of two-mode squeezed vacuum state under one-mode and two-mode laser channel respectively. We find the gain process will introduce additional noise to the two-mode squeezed vacuum state such that the steerable time is reduced. Secondly, by quantising quantum Einstein-Podolsky-Rosen steering, it shows that two-side loss presents a smaller steerability than one-side loss  although they share the same two-way steerable time.  In addition, we find the more gained party can steer the other's state, while the other party cannot steer the gained party in a certain threshold value. In this sense, it seems that the gain effect in one party is equivalent to the loss effect in the other’s party. Our results pave way for the distillation of Einstein-Podolsky-Rosen steering and the quantum information processing in practical quantum channels.

\end{abstract}
\maketitle

\section{Introduction}
The Einstein-Podolsky-Rosen (EPR) steering was first proposed by Schrödinger in 1935 in response to the famous paper by Einstein, Podolsky and Rosen on the completeness of quantum mechanics \cite{schrodinger1935discussion}. It allows the subsystems Alice steer the quantum state of a distant subsystem Bob during a local measurement \cite{reid2009colloquium,gallego2015resource,he2015classifying,cavalcanti2016quantum,uola2020quantum}. In terms of the degree of quantum correlation, EPR steering is intermediate between Bell inequality and quantum entanglement \cite{wiseman2007steering}. Unlike Bell inequality and quantum entanglement, EPR steering is inherently asymmetric for two subsystems, i.e., the ability of Alice to steer Bob may be different from the ability of Bob to steer Alice, and there may even be a one-way EPR steering situation \cite{armstrong2015multipartite,sun2016experimental,deng2017demonstration,tischler2018conclusive,cavailles2018demonstration,weston2018heralded}. EPR steering has received increasing attention since Wiseman et al. have analysed quantum entanglement, EPR steering and Bell inequality separately from the perspective of quantum information tasks \cite{wiseman2007steering}.
   
One important element of research on quantum EPR steering is to demonstrate the existence of one-way steering. In terms of discrete-variable systems, the one-way EPR steering was first demonstrated with Werner states and have been proved by using the "all-or-nothing proof" criterion and quantum storage \cite{saunders2010experimental,sun2014experimental}. Pan et al. have demonstrated the true four-body EPR steering and its application to one-way quantum computing \cite{li2015genuine}. For the continuously variable systems, the existence of one-way EPR steering based on Reid's criterion was first demonstrated by introducing the loss of one-sided paths in two-mode squeezed state \cite{handchen2012observation}.
On this basis, Su et al. have successfully realized the free control of EPR steering by using noise, and the many-body Gaussian EPR steering have been observed by introducing asymmetric losses \cite{qin2017manipulating}. On the other hand, the interaction between quantum system and environment is inevitable for the quantum information processing in practical quantum channels. Thus it is crucial to investigate the effect of decoherence on EPR steering in quantum channels, which can help to design effective distillation programmes. The decoherence of EPR steering have been theoretical analysed and experimentally demonstrated in recent years  \cite{rosales2015decoherence,deng2021sudden,wang2022stable,liu2022distillation}. However, most of them only take account in lossy and noisy channels. To avoid the decoherence of EPR steering and accomplish the quantum information processing in practical quantum channels, we should
analyse the evolution of EPR steering in more general environments, such as laser channel and phase-sensitive environment.

We first derive the relation between the probability distribution functions and the characteristic function in Sec. II. The relation provide a relative easy calculation for the analytical expressions of EPR steering criterion for two-mode squeezed vacuum state.  In Sec. III, we analyse the steering conditions of two-mode squeezed vacuum state under laser channel. In Sec. IV, we further analyse the relationship between magnitude of EPR steering and the parameters of the laser channel.
In Sec. V, we make a comparison about the decoherence between quantum steering and
entanglement in laser channel. In Sec. VI, we extend our discussions to the evolution of the TMSV in the presence of phase-sensitive environment. Lastly, in Sec. VII, we give concluding remarks.

\bigskip

\section{The EPR steering criterion}

\subsection{Relation between PDFs and CF}

In quantum steering criterion, the probability distribution functions (PDFs)
are usually used in the calculation of original Reid criterion, entropic steering criterion, and Adesso's steering low bound for Gaussian states. In addition, it is worth noting that the characteristic function (CF) provides a relative easy
calculation for the analytical expressions of the average values, especially
for the systems in noisy environments. Here, thus we first derive the relation
between PDFs and CF.

For a two-mode quantum system $\rho$, its Weyl expression of density operator
is given by
\begin{align}
\rho &  =\int_{-\infty}^{\infty}\frac{dq_{1}dp_{1}dq_{2}dp_{2}}{\pi^{2}}%
\chi \left(  q_{1},p_{1};q_{2},p_{2}\right) \nonumber \\
&  \times D_{1}\left(  -q_{1},-p_{1}\right)  D_{2}\left(  -q_{2}%
,-p_{2}\right)  , \label{s1}%
\end{align}
where $\chi \left(  q_{1},p_{1};q_{2},p_{2}\right)  =\mathtt{tr}\left[  \rho
D_{1}\left(  q_{1},p_{1}\right)  D_{2}\left(  q_{2},p_{2}\right)  \right]  $
is the CF and $D_{l}\left(  q_{l},p_{l}\right)  $ ($l=1,2$) are the
displacement operators. In addition, $D_{l}\left(  q_{l},p_{l}\right)  $ can
also be shown as%
\begin{align}
D_{l}\left(  q_{l},p_{l}\right)   &  =e^{i(p_{l}Q_{l}-q_{l}P_{l})}\nonumber \\
&  =e^{-iq_{l}p_{l}}e^{ip_{l}Q_{l}}e^{-iq_{l}P_{l}}\label{s2}\\
&  =e^{iq_{l}p_{l}}e^{-iq_{l}P_{l}}e^{ip_{l}Q_{l}}, \label{s3}%
\end{align}
where $Q_{l}=a_{l}+a_{l}^{\dag}$ and $P_{l}=(a_{l}-a_{l}^{\dag})/i$ are the
coordinate and momentum operators, respectively, satisfying $\left[
Q_{l},P_{l}\right]  =2i$. From Eqs. (\ref{s2}) and (\ref{s3}) and noticing
that $e^{-iq_{l}P_{l}}\left \vert \bar{q}_{l}\right \rangle =\left \vert \bar
{q}_{l}+\sqrt{2}q_{l}\right \rangle $ as well as $e^{ip_{l}Q_{l}}\left \vert
\bar{p}_{l}\right \rangle =\left \vert \bar{p}_{l}+\sqrt{2}p_{l}\right \rangle $,
it is ready to see%
\begin{align}
\left \langle \bar{q}_{l}\right \vert D_{l}\left(  q_{l},p_{l}\right)
\left \vert \bar{q}_{l}\right \rangle  &  =e^{i\sqrt{2}\bar{q}_{l}p_{l}}%
\delta \left(  \sqrt{2}q_{l}\right)  ,\label{s4}\\
\left \langle \bar{p}_{l}\right \vert D_{l}\left(  q_{l},p_{l}\right)
\left \vert \bar{p}_{l}\right \rangle  &  =e^{-i\sqrt{2}q_{l}\bar{p}_{l}}%
\delta \left(  \sqrt{2}p_{l}\right)  , \label{s5}%
\end{align}
where $\left \vert \bar{q}_{l}\right \rangle $ and $\left \vert \bar{p}%
_{l}\right \rangle $ are respectively the eigenstates of $Q_{l}$ and $P_{l},$
i.e., $Q_{l}\left \vert \bar{q}_{l}\right \rangle =\sqrt{2}\bar{q}_{l}\left \vert
\bar{q}_{l}\right \rangle $, $P_{l}\left \vert \bar{p}_{l}\right \rangle
=\sqrt{2}\bar{p}_{l}\left \vert \bar{p}_{l}\right \rangle $.

After substituting Eqs.(\ref{s4}) and (\ref{s5}) into Eq.(\ref{s1}), then the
joint PDFs and its marginal and conditional PDFs can be easily calculated by,
for instance,%
\begin{align}
P\left(\bar{q}_{1},\bar{q}_{2}\right)   &  =\left \langle \bar{q}_{1},\bar
{q}_{2}\right \vert \rho \left \vert \bar{q}_{1},\bar{q}_{2}\right \rangle
\nonumber \\
&  =\int_{-\infty}^{\infty}\frac{dp_{1}dp_{2}}{2\pi^{2}}e^{-i\sqrt{2}\left(
\bar{q}_{1}p_{1}+\bar{q}_{2}p_{2}\right)  }
 \chi \left(  0,p_{1};0,p_{2}\right)  , \label{s6}%
\end{align}%
\begin{align*}
P\left(  \bar{p}_{1},\bar{p}_{2}\right)   &  =\left \langle \bar{p}_{1},\bar
{p}_{2}\right \vert \rho \left \vert \bar{p}_{1},\bar{p}_{2}\right \rangle \\
&  =\int_{-\infty}^{\infty}\frac{dq_{1}dq_{2}}{2\pi^{2}}e^{i\sqrt{2}\left(
\bar{p}_{1}q_{1}+\bar{p}_{2}q_{2}\right)  }\chi \left(  q_{1},0;q_{2},0\right)  ,
\end{align*}
and%
\begin{align}
P\left(  \bar{q}_{1}\right)   &  =\int_{-\infty}^{\infty}d\bar{q}_{2}P\left(
\bar{q}_{1},\bar{q}_{2}\right) \nonumber \\
&  =\frac{1}{\sqrt{2}\pi}\int_{-\infty}^{\infty}dp_{1}e^{-i\sqrt{2}\bar{q}%
_{1}p_{1}}\chi \left(  0,p_{1};0,0\right)  , \label{s7}%
\end{align}%
\begin{align*}
P\left(  \bar{p}_{1}\right)   &  =\int_{-\infty}^{\infty}d\bar{p}_{2}P\left(
\bar{p}_{1},\bar{p}_{2}\right) \\
&  =\frac{1}{\sqrt{2}\pi}\int_{-\infty}^{\infty}dq_{1}e^{i\sqrt{2}\bar{p}%
_{1}q_{1}}\chi \left(  q_{1},0;0,0\right)  ,
\end{align*}
as well as%
\begin{equation}
P\left(  \left.  \bar{q}_{2}\right \vert \bar{q}_{1}\right)  =P\left(  \bar
{q}_{1},\bar{q}_{2}\right)  /P\left(  \bar{q}_{1}\right)  . \label{s8}%
\end{equation}
Thus the inferred variance of $q$ can be derived by%
\begin{align}
&  \Delta_{\inf}^{2}\left(  \left.  Q_{2}\right \vert Q_{1}\right) \nonumber \\
&  \equiv \int_{-\infty}^{\infty}d\bar{q}_{1}P\left(  \bar{q}_{1}\right)
\Delta_{\inf}^{2}\left(  \left.  Q_{2}\right \vert Q_{1}=\bar{q}_{1}\right)
\nonumber \\
&  =\int_{-\infty}^{\infty}d\bar{q}_{1}P\left(  \bar{q}_{1}\right)
\int_{-\infty}^{\infty}d\bar{q}_{2}P\left(  \left.  \bar{q}_{2}\right \vert
\bar{q}_{1}\right)  \left(  \bar{q}_{2}-q_{est}\right)  ^{2}\nonumber \\
&  =\int_{-\infty}^{\infty}d\bar{q}_{1}d\bar{q}_{2}P\left(  \bar{q}_{1}%
,\bar{q}_{2}\right)  \left(  \bar{q}_{2}-q_{est}\right)  ^{2}. \label{s9}%
\end{align}
Here we should note that $q_{est}$ is the function of variable $\bar{q}_{1}$.

\subsection{Reid criterion}

Using the linear inferences, i.e., taking $q_{est}=d-\lambda \bar{q}_{1}$,
where the two parameters $\lambda$ and $d$ will be chosen to minimize the
inferred variance $\Delta_{\inf}^{2}\left(  \left.  Q_{2}\right \vert
Q_{1}\right)  $, after a straight calculation then one can get%
\begin{align}
\Delta_{\inf}^{2}\left(  \left.  Q_{2}\right \vert Q_{1}\right)   &  =\frac
{1}{2}\lambda^{2}\left \langle Q_{1}^{2}\right \rangle +\lambda \left \langle
Q_{1}Q_{2}\right \rangle -\sqrt{2}\lambda d\left \langle Q_{1}\right \rangle
\nonumber \\
&  +d^{2}-\sqrt{2}d\left \langle Q_{2}\right \rangle +\frac{1}{2}\left \langle
Q_{2}^{2}\right \rangle , \label{s10}%
\end{align}
where we have used the relation between the CF and the average values, for
instance,%
\begin{align}
\left \langle Q_{2}\right \rangle  &  =\left.  \frac{\partial}{\partial \left(
ip_{2}\right)  }\chi \left(  0,0;0,p_{2}\right)  \right \vert _{p_{2}%
=0},\label{s11}\\
\left \langle Q_{2}^{2}\right \rangle  &  =\left.  \frac{\partial^{2}}%
{\partial \left(  ip_{2}\right)  ^{2}}\chi \left(  0,0;0,p_{2}\right)
\right \vert _{p_{2}=0},\label{s12}\\
\left \langle Q_{1}Q_{2}\right \rangle  &  =\left.  \frac{-\partial^{2}%
}{\partial p_{1}\partial p_{2}}\chi \left(  0,p_{1};0,p_{2}\right)  \right \vert
_{p_{1},p_{2}=0}. \label{s13}%
\end{align}
The values of the two parameters $\lambda$ and $d$ are chosen by minimizing
$\Delta_{\inf}^{2}\left(  \left.  Q_{2}\right \vert Q_{1}\right)  $ such that
$\frac{\partial}{\partial \lambda}\Delta_{\inf}^{2}\left(  \left.
Q_{2}\right \vert Q_{1}\right)  =\frac{\partial}{\partial d}\Delta_{\inf}%
^{2}\left(  \left.  Q_{2}\right \vert Q_{1}\right)  =0$, from which it follows
that%
\begin{equation}
\lambda=-\frac{E_{Q_{1}Q_{2}}}{V\left(  Q_{1}\right)  },d=\frac{\lambda
\left \langle Q_{1}\right \rangle +\left \langle Q_{2}\right \rangle }{\sqrt{2}},
\label{s14}%
\end{equation}
where $E_{Q_{1}Q_{2}}=\left \langle Q_{1}Q_{2}\right \rangle -\left \langle
Q_{1}\right \rangle \left \langle Q_{2}\right \rangle $ and $V\left(
Q_{l}\right)  =\left \langle Q_{l}^{2}\right \rangle -\left \langle
Q_{l}\right \rangle ^{2}$ is the voriance for $Q_{l}$, $l=1,2$. Substituting
Eq. (\ref{s14}) into Eq. (\ref{s10}), the minimized inferred variance
$\Delta_{\min}^{2}\left(  \left.  Q_{2}\right \vert Q_{1}\right)  $ is given
by
\begin{equation}
\Delta_{\min}^{2}\left(  \left.  Q_{2}\right \vert Q_{1}\right)  =\frac{1}%
{2}\left[  V\left(  Q_{2}\right)  -\frac{\left(  E_{Q_{1}Q_{2}}\right)  ^{2}%
}{V\left(  Q_{1}\right)  }\right]  , \label{s15}%
\end{equation}
The minimized inferred variance $\Delta_{\min}^{2}\left(  \left.
P_{2}\right \vert P_{1}\right)  $ can be similarly derived as%
\begin{equation}
\Delta_{\min}^{2}\left(  \left.  P_{2}\right \vert P_{1}\right)  =\frac{1}%
{2}\left[  V\left(  P_{2}\right)  -\frac{\left(  E_{P_{1}P_{2}}\right)  ^{2}%
}{V\left(  P_{1}\right)  }\right]  , \label{s16}%
\end{equation}
where $V\left(  P_{l}\right)  =\left \langle P_{l}^{2}\right \rangle
-\left \langle P_{l}\right \rangle ^{2}$ and $E_{P_{1}P_{2}}=\left \langle
P_{1}P_{2}\right \rangle -\left \langle P_{1}\right \rangle \left \langle
P_{2}\right \rangle $. Combing Eqs. (\ref{s15}) and (\ref{s16}) and employing
the Heisenberg uncertainty relation, yields the Reid inequality (RI)%
\begin{equation}
V_{B|A}\equiv \Delta_{\min}^{2}\left(  \left.  Q_{2}\right \vert Q_{1}\right)
\Delta_{\min}^{2}\left(  \left.  P_{2}\right \vert P_{1}\right)  \geqslant
\frac{1}{4}, \label{s17}%
\end{equation}
where the communtation relations $\left[  Q_{l},P_{l}\right]  =2i$ are used.
Thus the EPR paradox---or steering occurs when the RI is violated.

In a similar way, if Bob wants to infer Alice's measurement results, then the
RI is
\begin{equation}
V_{A|B}\equiv \Delta_{\min}^{2}\left(  \left.  Q_{1}\right \vert Q_{2}\right)
\Delta_{\min}^{2}\left(  \left.  P_{1}\right \vert P_{2}\right)  \geqslant
\frac{1}{4}, \label{s18}%
\end{equation}
with
\begin{align}
\Delta_{\min}^{2}\left(  \left.  Q_{1}\right \vert Q_{2}\right)   &  =\frac
{1}{2}\left[  V\left(  Q_{1}\right)  -\frac{\left(  E_{Q_{1}Q_{2}}\right)
^{2}}{V\left(  Q_{2}\right)  }\right]  ,\label{s19}\\
\Delta_{\min}^{2}\left(  \left.  P_{1}\right \vert P_{2}\right)   &  =\frac
{1}{2}\left[  V\left(  P_{1}\right)  -\frac{\left(  E_{P_{1}P_{2}}\right)
^{2}}{V\left(  P_{2}\right)  }\right]  . \label{s20}%
\end{align}
From Eqs. (\ref{s15})-(\ref{s20}), it is clear that when the CF is known, then
$V_{B|A}$ and $V_{A|B}$ can be directly calculated using Eqs. (\ref{s11}%
)-(\ref{s13}). In addition, for Guassian states which is fully described by
its covariance matrix, these elements involved in the inferred variance can be
directly extracted from the covariance matrix.

\subsection{RI violation of two-mode squeezed vacuum state}

Next, let us consider a two-mode squeezed vacuum (TMSV) state $\left \vert
TMSV\right \rangle $, which can be generated by a nondegenerate optical
parameter process, i.e.,
\begin{equation}
\left \vert TMSV\right \rangle =S\left(  r\right)  \left \vert 00\right \rangle ,
\label{s21}%
\end{equation}
where $S\left(  r\right)  =\exp \{r(a_{1}^{\dag}a_{2}^{\dag}-a_{1}a_{2})\}$ is
the two-mode squeezing operator. The corresponding CF can be calculated as
\begin{align}
\chi_{TMSV}  &  =\exp \left \{  -\frac{1}{2}(q_{1}^{2}+q_{2}^{2}+p_{1}^{2}%
+p_{2}^{2})\cosh2r\right \} \nonumber \\
&  \times \exp \left \{  (q_{1}q_{2}-p_{1}p_{2})\sinh2r\right \}  . \label{s22}%
\end{align}%
Using Eqs. (\ref{s11})-(\ref{s13}), it is ready to have
\begin{equation}
\Delta_{\min}^{2}\left(  \left.  Q_{2}\right \vert Q_{1}\right)  =\Delta_{\min
}^{2}\left(  \left.  P_{2}\right \vert P_{1}\right)  =\frac{1}{2\cosh2r},
\label{s23}%
\end{equation}
which leads to
\begin{equation}
V_{B|A}=\frac{1}{4\cosh^{2}2r}<\frac{1}{4}. \label{s24}%
\end{equation}
Thus the TMSV shows EPR steering for any nonzero squeezing.$\allowbreak$ Here
we should mention that Reid criterion actually depends on the second-order
moments of observables, which implies that Reid criterion is effective for
Gaussian states because Gaussian states can be fully described by second-order
moments or covariance matrix. However, this case is not true for non-Gaussian
states which also depends on its higher order moments.

\subsection{Entropic steering criterion}

To more effectively demonstrate the EPR steering especially for non-Gaussian
states, correlations in all orders should be involved for steering criterion.
Based on the entropic uncertainty relation \cite{bialynicki1975uncertainty},
\begin{equation}
H\left(  Q\right)  +H\left(  P\right)  \geqslant \ln \left(  e\pi \right)  ,
\label{s25}%
\end{equation}
where $H\left(  Y\right)  $ ($Y=Q,P$ representing observables) is the Shannon
entropy for $y\left(  =q,p\right)  $ PDF, i.e.,
\begin{equation}
H\left(  Y\right)  =-\int dyP\left(  y\right)  \ln P\left(  y\right)  ,
\label{s26}%
\end{equation}
Walborn \textit{et al}. have derived an entropic
steering inequality \cite{walborn2011revealing}, i.e., for instance, for the position and momentum
distribution,%
\begin{equation}
H_{B|A}\equiv H\left(  \left.  Q_{2}\right \vert Q_{1}\right)  +H\left(
\left.  P_{2}\right \vert P_{1}\right)  \geqslant \ln \left(  e\pi \right)  ,
\label{s27}%
\end{equation}
where the conditional entropies $H\left(  \left.  Q_{2}\right \vert
Q_{1}\right)  $ and $H\left(  \left.  P_{2}\right \vert P_{1}\right)  $ are
respectively given by%
\begin{align}
H\left(  \left.  Q_{2}\right \vert Q_{1}\right)   &  =H\left(  Q_{1}%
,Q_{2}\right)  -H\left(  Q_{1}\right)  ,\label{s28}\\
H\left(  \left.  P_{2}\right \vert P_{1}\right)   &  =H\left(  P_{1}%
,P_{2}\right)  -H\left(  P_{1}\right)  , \label{s29}%
\end{align}
with%
\begin{align}
H\left(  x_{1}\right)   &  =-\int dx_{1}P\left(  x_{1}\right)  \ln P\left(
x_{1}\right)  ,\label{s30}\\
H\left(  x_{1},x_{2}\right)   &  =-\int dx_{1}dx_{2}P\left(  x_{1}%
,x_{2}\right)  \ln P\left(  x_{1},x_{2}\right)  . \label{s31}%
\end{align}

Next, we consider the inferred Shannon entropy $H_{B|A}$ for the TMSV shown in
Eq. (\ref{s21}) as an example. Using the CF of the TMSV in Eq. (\ref{s22}) and
Eqs. (\ref{s6}) and (\ref{s7}), it is ready to have
\begin{align}
P\left(  \bar{q}_{1}\right)   &  =\frac{1}{\sqrt{\pi \cosh2r}}e^{-\bar{q}%
_{1}^{2}\operatorname{sech}2r},\label{s32}\\
P\left(  \bar{p}_{1}\right)   &  =\frac{1}{\sqrt{\pi \cosh2r}}e^{-\bar{p}%
_{1}^{2}\operatorname{sech}2r}, \label{s33}%
\end{align}
and
\begin{align}
P\left(  \bar{q}_{1},\bar{q}_{2}\right)   &  =\frac{1}{\pi}e^{-(\bar{q}%
_{1}^{2}+\bar{q}_{2}^{2})\cosh2r+2\bar{q}_{1}\bar{q}_{2}\sinh2r},\label{s34}\\
P\left(  \bar{p}_{1},\bar{p}_{2}\right)   &  =\frac{1}{\pi}e^{-(\bar{p}%
_{1}^{2}+\bar{p}_{2}^{2})\cosh2r-2\bar{p}_{1}\bar{p}_{2}\sinh2r}, \label{s35}%
\end{align}
then we have%
\begin{align}
H\left(  Q_{1}\right)   &  =H\left(  P_{1}\right)  =\ln \sqrt{\pi e\cosh
2r}\label{s36}\\
H\left(  Q_{1},Q_{2}\right)   &  =H\left(  P_{1},P_{2}\right)  =\ln e\pi,
\label{s37}%
\end{align}
and%
\begin{equation}
H\left(  \left.  Q_{2}\right \vert Q_{1}\right)  =H\left(  \left.
P_{2}\right \vert P_{1}\right)  =\frac{1}{2}\ln \frac{e\pi}{\cosh2r},
\label{s38}%
\end{equation}
as well as%
\begin{equation}
H\left(  \left.  Q_{2}\right \vert Q_{1}\right)  +H\left(  \left.
P_{2}\right \vert P_{1}\right)  =\ln \frac{e\pi}{\cosh2r}<\ln e\pi, \label{s39}%
\end{equation}
where we have used the following derived formula%
\begin{equation}
P\left(  x\right)  =\sqrt{\frac{1}{\pi a}}e^{-\frac{1}{a}x^{2}},H\left(
X\right)  =\ln \sqrt{\pi ae}, \label{s40}%
\end{equation}
and
\begin{align}
P\left(  x_{1},x_{2}\right)   &  =\frac{\sqrt{a^{2}-b^{2}}}{\pi}%
e^{-a(x_{1}^{2}+x_{2}^{2})+2x_{1}x_{2}b},\label{s41}\\
H\left(  X_{1},X_{2}\right)   &  =\ln \frac{e\pi}{\sqrt{a^{2}-b^{2}}},a>b.
\label{s42}%
\end{align}
Comparing Eq. (\ref{s39}) with Eq. (\ref{s24}), it is ready to see that Reid
and entropic criteria lead to the same result for the TMSV. Actually, for the
Gaussian states, both the entropic and Heisenberg uncertainty relation is
equivalent, thus the same results can be shared by both of them for
position-momentum measurement \cite{wolf2006extremality}. Here we should mention
that for non-Gaussian states, the case is not always true. Generally, the
entropic steering criterion performs better than the Reid one for non-Gaussian
states \cite{lee2013quantum,PhysRevA.89.012104}.

\section{EPR Steering under laser channel}

In this section, we consider the quantum steering of quantum states with
continuous variable, such as the TMSVs, when both modes are going
through laser channel which includes both gain and loss effects. For this
purpose, here we start with the master equation and establish the output-input
relation of the CF.

\subsection{Laser channel and the output-input CF relation}

For single-mode system, the master equation describing the laser channel is
given by \cite{da2019photocounting}%
\begin{align}
\frac{d}{dt}\rho \left(  t\right)   &  =g\left[  2a^{\dag}\rho a-aa^{\dag}%
\rho-\rho aa^{\dag}\right] \nonumber \\
&  +\kappa \left[  2a\rho a^{\dag}-a^{\dag}a\rho-\rho a^{\dag}a\right]  ,
\label{s43}%
\end{align}
where $g$ and $\kappa$ are the gain and the loss factors, respectively. In
particular, when $g=0$ then Eq. (\ref{s43}) reduces to the master equation
describing the amplitude decay channel (photon-loss channel); while for the
case with $g\rightarrow \kappa \bar{n}$ and $\kappa \rightarrow \kappa \left(
\bar{n}+1\right)  ,$ then Eq. (\ref{s43}) becomes the master equation
describing the thermal channel \cite{gardiner2000quantum}.

To establish the output-input relation, here we can employ the entangled state
representation to obtain (see Appendix A )%
\begin{equation}
\chi_{out}\left(  \alpha,\alpha^{\ast}\right)  =e^{-\frac{A}{2}|\alpha|^{2}%
}\chi_{in}\left(  \alpha e^{-\left(  \kappa-g\right)  t},\alpha^{\ast
}e^{-\left(  \kappa-g\right)  t}\right)  , \label{s44}%
\end{equation}
where
\begin{equation}
A=\Omega \left(  1-R\right)  ,\Omega=\frac{\kappa+g}{\kappa-g},R=e^{-2\left(
\kappa-g\right)  t}. \label{s45}%
\end{equation}
Eq. (\ref{s44}) is just the output-input relation for any single-mode quantum
state through the laser channel. When both modes of two-mode quantum states
independently go through the laser channel, from Eq. (\ref{s44}) it is ready
to have
\begin{align}
&  \chi_{out}\left(  \alpha,\alpha^{\ast};\beta,\beta^{\ast}\right)
\nonumber \\
&  =e^{-\frac{A}{2}(|\alpha|^{2}+|\beta|^{2})}\chi_{in}\left(  \tilde{\alpha
},\tilde{\alpha}^{\ast};\tilde{\beta},\tilde{\beta}^{\ast}\right)
,\label{s46}\\
&  \left.  \left(  \tilde{\alpha}=\alpha e^{-\left(  \kappa-g\right)
t},\tilde{\beta}=\beta e^{-\left(  \kappa-g\right)  t}\right)  \right.
,\nonumber
\end{align}
where for simplicity, the symmetrical laser channels are assumed, which are
characteristics of same gain and loss factors for two-mode.

\subsection{TMSV under two-mode laser channel}

Here, we first consider the effects of symmetrical laser channels on the
steerability of the decohered TMSV. Combining Eqs. (\ref{s22}) and
(\ref{s46}), the output CF of the TMSV under the laser channel can be readily
gotten
\begin{align}
&  \chi_{out}\left(  q_{1},p_{1};q_{2},p_{2}\right) \nonumber \\
&  =\exp \left \{  -\frac{B}{2}(q_{1}^{2}+q_{2}^{2}+p_{1}^{2}+p_{2}^{2})\right \}
\nonumber \\
&  \times \exp \left \{  C(q_{1}q_{2}-p_{1}p_{2})\right \}  . \label{s47}%
\end{align}
where we have taken $\alpha=q_{1}+ip_{1}$, $\beta=q_{2}+ip_{2},$ and
\begin{equation}
B=A+R\cosh2r,C=R\sinh2r. \label{s48}%
\end{equation}
Thus, using Eqs. (\ref{s6}), (\ref{s7}) and (\ref{s47}), the joint position
and its marginal PDFs can be given by%
\begin{align}
P\left(  \bar{q}_{1}\right)   &  =\frac{1}{\sqrt{\pi B}}\exp \left \{
-\frac{\bar{q}_{1}^{2}}{B}\right \}  ,\label{s50}\\
P\left(  \bar{q}_{1},\bar{q}_{2}\right)   &  =\frac{1}{\pi \sqrt{B^{2}-C^{2}}%
}e^{-\frac{B\left(  \bar{q}_{1}^{2}+\bar{q}_{2}^{2}\right)  -2C\bar{q}_{1}%
\bar{q}_{2}}{B^{2}-C^{2}}}. \label{s51}%
\end{align}
The joint momentum and its marginal PDFs have the same functional form as Eqs.
(\ref{s51}) and (\ref{s50}), respectively, just making a replacement with
$\bar{q}_{1},\bar{q}_{2}$ by $\bar{p}_{1},-\bar{p}_{2}.$ From Eqs. (\ref{s51})
and (\ref{s50}), it is ready to have%
\begin{equation}
H\left(  \left.  Q_{2}\right \vert Q_{1}\right)  =H\left(  \left.
P_{2}\right \vert P_{1}\right)  =\frac{1}{2}\ln \frac{\pi e(B^{2}-C^{2})}{B},
\label{s52}%
\end{equation}
which leads to
\begin{equation}
H\left(  \left.  Q_{2}\right \vert Q_{1}\right)  +H\left(  \left.
P_{2}\right \vert P_{1}\right)  =\ln \frac{\pi e(B^{2}-C^{2})}{B}. \label{s53}%
\end{equation}
On the other hand, from Eq. (\ref{s47}) and using Eqs. (\ref{s11}%
)-(\ref{s13}), the inferred variances are derived as%
\begin{equation}
\Delta_{\min}^{2}\left(  \left.  Q_{2}\right \vert Q_{1}\right)  =\Delta_{\min
}^{2}\left(  \left.  P_{2}\right \vert P_{1}\right)  =\frac{B^{2}-C^{2}}{2B},
\label{s54}%
\end{equation}
and consequently the steerable condition is given by%
\begin{equation}
\Delta_{\min}\left(  \left.  Q_{2}\right \vert Q_{1}\right)  \Delta_{\min
}\left(  \left.  P_{2}\right \vert P_{1}\right)  =\frac{B^{2}-C^{2}}{2B}.
\label{s55}%
\end{equation}
Comparing Reid and entropic steering conditions by Eqs. (\ref{s53}) and
(\ref{s55}), the same steerable condition is shared, as expected, i.e.,
\begin{equation}
B^{2}-C^{2}<B. \label{s56}%
\end{equation}

Substituting Eqs. (\ref{s45}) and (\ref{s48}) into Eq. (\ref{s56}) and
noticing $A=\Omega(1-R)>0$, we can rewritte the steerable condition as%
\begin{equation}
aA^{2}+bA-c<0, \label{s57}%
\end{equation}
where we have set%

\begin{align*}
a  &  =\frac{1}{\Omega^{2}}\left(  \Omega^{2}+1-2\Omega \cosh2r\right)  ,\\
b  &  =\frac{1}{\Omega}\left(  2\Omega \cosh2r+\cosh2r-2-\Omega \right)  ,\\
c  &  =\cosh2r-1=2\sinh^{2}r>0.
\end{align*}
After a straight discussion, it shown that the steerable condition of two-way is
\begin{equation}
t<t_{c}=\frac{1}{2\left(  \kappa-g\right)  }\ln \left \{  1+\frac{b-\sqrt
{b^{2}+4ac}}{2\Omega a}\right \}  ^{-1}. \label{s57b}%
\end{equation}
 Here we obtained
a general expression (\ref{s57b}) that applies to linear amplifiers as well.
In generally, the upper bound of evolution time $t_{c}$ not only depends on
the gain and loss factors, but also on the squeezing parameter. The
photon-loss, photon-gain and thermal noise channels can be seen as three
special case. In particular, for the photon-loss channel, $g=0$, then the
steerable condition is given by%
\begin{equation}
\kappa t<\frac{1}{2}\ln2. \label{s58}%
\end{equation}
It is interesting to notice that this steerable condition is independent of
the TMSV only with $r>0.$ From Eq. (\ref{s58}), the decohered TMSV is
steerable provided the photon-loss is less than 50\%, i.e., $e^{-2\kappa
t}>1/2$. While for the thermal channel, $g\rightarrow \kappa \bar{n}$ and
$\kappa \rightarrow \kappa \left(  \bar{n}+1\right)  ,$ thus when $0<\bar
{n}<\left(  e^{2r}-1\right)  /2,$ then the steerable condition is
\begin{equation}
\kappa t<\frac{1}{2}\ln \frac{2\left \vert \alpha \right \vert }{\beta+\sqrt
{\beta^{2}+4\left \vert \alpha \right \vert \delta}}, \label{s59}%
\end{equation}
where $\alpha=(N+1)^{2}-4N\cosh^{2}r$, $\beta=\left(  2N-1\right)  \left(
\cosh2r-N\right)  $, $\delta=N(N-1)>0,$ and $N=\left(  2\bar{n}+1\right)  >1$.
In particular, when $\bar{n}=0,$ Eq. (\ref{s59}) just reduces to Eq.
(\ref{s58}), as expected.

At the end of this section, we consider the case with $\kappa=0$, leading to
$\Omega=-1,$ then from Eq. (\ref{s57b}) we have%
\begin{equation}
gt<\frac{1}{2}\ln \frac{3+\sqrt{1+8\tanh^{2}r}}{4}\leqslant \frac{1}{2}\ln
\frac{3}{2}. \label{s60}%
\end{equation}
From Eq. (\ref{s60}) it is clear that when the gain $g$ exists but no loss,
the steerable condition depends on the squeezing parameter $r$, which is
different from the case for photon-loss in Eq. (\ref{s58}). It is interesting
to notice that, however, for any squeezing parameter, the steerable condition
will no longer be satisfied if the gain exceeds a threshold, i.e.,
$gt>\frac{1}{2}\ln \frac{3}{2}$. This indicates that the gain will introduce
additional noise to the TMSV such that the steerable condition is destroyed.
In addition, comparing Eqs. (\ref{s60}) with (\ref{s58}), it is shown that the
gain process will provide a shorter steerable time if taking $g=\kappa,$ the
limit difference between both is $\frac{1}{2}\ln \frac{4}{3}$.

\begin{figure}[ptb]
\label{F1-1} \centering \includegraphics[width=8cm]{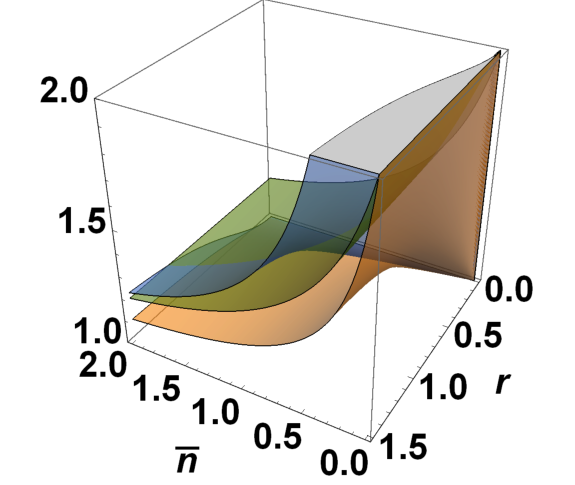} \newline%
\caption{{}(Color online) Comparison of steerable time as a function of
$\bar{n}$ and $r$ for one- and two-side thermal channel. The vertical coordinate is $e^{2\kappa t}$. Blue (green) and
orange surfaces correspond to one-side and two-side cases, respectively.}%
\end{figure}

\subsection{One way steering}

In fact, the quantum steering is characteristized by asymmetry, i.e., system A
can steer system B, but the case is not true vice verse. In order to clearly
show the asymmetry of steering, we examine the effect of gain and loss on the
steering parameters with respect to system A and system B when the TMSV is
under one-mode laser channel.
Let us assume that system B (mode $b$) is coupled to the laser channel. In
this case, according to Eq. (\ref{s46}), the CF of the final state is given by%
\begin{equation}
\chi_{out}\left(  \alpha,\alpha^{\ast};\beta,\beta^{\ast}\right)
=e^{-\frac{A}{2}|\beta|^{2}}\chi_{in}\left(  \alpha,\alpha^{\ast};\tilde
{\beta},\tilde{\beta}^{\ast}\right)  , \label{s61}%
\end{equation}
and the output CF of the TMSV under the one-side laser channel is%
\begin{align}
\chi_{TMSV}  &  =\exp \left \{  -\frac{\cosh2r}{2}(q_{1}^{2}+p_{1}^{2})\right \}
\nonumber \\
&  \times \exp \left \{  -\frac{B}{2}(q_{2}^{2}+p_{2}^{2})\right \} \nonumber \\
&  \times \exp \left \{  \frac{C}{\sqrt{R}}(q_{1}q_{2}-p_{1}p_{2})\right \}  .
\label{s62}%
\end{align}
From Eq. (\ref{s62}) it is ready to have the covariance matrix $\Sigma$,
i.e.,
\begin{equation}
\Sigma=\left(
\begin{array}
[c]{cccc}%
\cosh2r & 0 & \frac{C}{\sqrt{R}} & 0\\
0 & \cosh2r & 0 & -\frac{C}{\sqrt{R}}\\
\frac{C}{\sqrt{R}} & 0 & B & 0\\
0 & -\frac{C}{\sqrt{R}} & 0 & B
\end{array}
\right)  , \label{s63}%
\end{equation}
which becomes the covariance matrix of the TMSV under ideal condition of
$t=0$, as expected. Using Eq. (\ref{s63}) and Eqs. (\ref{s15})-(\ref{s20}), it
is ready to have the steering of $B$ $\left(  \text{mode }b\right)  $ and of
$A$ $\left(  \text{mode }a\right)  $, respectively,
\begin{align}
V_{B|A}  &  =\frac{1}{4}\left(  A+\frac{R}{\cosh2r}\right)  ^{2},\label{s64}\\
V_{A|B}  &  =\frac{1}{4}\left(  \frac{R+A\cosh2r}{A+R\cosh2r}\right)  ^{2}.
\label{s65}%
\end{align}
In particular, when $R=1$ ($t=0$) $V_{B|A}=V_{A|B}=1/(2\cosh2r)^{2},$ as expected.

For the steering of $B$ ($A$ steering $B$), using Eq. (\ref{s64}) and notice
$R>0$ and $A>0$, the steerable condition is obtained as%
\begin{equation}
t_{A\rightarrow B}<\frac{1}{2\left(  \kappa-g\right)  }\ln \frac{\kappa
\sinh^{2}r+g\cosh^{2}r}{g\cosh2r}, \label{s66}%
\end{equation}
Gerenally, the upbound time for the steering of $B$ depends on both loss
(gain) factor and squeeing parameter. In particular, there are three special
cases, i.e.,%
\begin{align}
\kappa t_{A\rightarrow B}  &  <\infty,\text{
\  \  \  \  \  \  \  \  \  \  \  \  \  \  \  \  \  \  \  \  \  \ loss,}\label{s67}\\
gt_{A\rightarrow B}  &  <\frac{1}{2}\ln \left(  2-\operatorname{sech}%
^{2}r\right)  \text{%
$<$%
}\frac{1}{2}\ln2,\text{ \ gain,}\label{s68}\\
\kappa t_{A\rightarrow B}  &  <\frac{1}{2}\ln \frac{N-\operatorname{sech}%
2r}{2\bar{n}},\text{thermal noise,} \label{s69}%
\end{align}
where $N=2\bar{n}+1$. Eqs. (\ref{s67})-(\ref{s69}) correspond to photon-loss
channel $\left(  g=0\right)  $, photon-gain channel $\left(  \kappa=0\right)
$ and thermal niose channel $\left(  g\rightarrow \kappa \bar{n},\kappa
\rightarrow \kappa \left(  \bar{n}+1\right)  \right)  $, respectively. It is
interesting that $A$ can always steer $B$ for any long decoherence time when
$B$ system passes only the loss channel. However, when $B$ system passes only
the gain channel, there exists a upbound $\frac{1}{2}\ln \left(
2-\operatorname{sech}^{2}r\right)  $ which is less than $\frac{1}{2}\ln2.$ In
addition, for the thermal noise channel, the upbound time depends on both
thermal average photon number $\bar{n}$ and squeezing parameter $r$.

In a similar way, for the steering of $A$ ($B$ steering $A$), we have%
\begin{equation}
t_{B\rightarrow A}<\frac{1}{2\left(  \kappa-g\right)  }\ln \frac{2\kappa
}{\kappa+g},\left(  r>0\right)  , \label{s70}%
\end{equation}
and
\begin{align}
\kappa t_{B\rightarrow A}  &  <\frac{1}{2}\ln2,\text{ \  \  \  \  \  \ loss,}%
\label{s71}\\
gt_{B\rightarrow A}  &  <\infty,\text{ \  \  \  \  \  \  \  \  \  \  \ gain,}%
\label{s72}\\
\kappa t_{B\rightarrow A}  &  <\frac{1}{2}\ln \frac{2\bar{n}+2}{2\bar{n}%
+1},\text{thermal noise.} \label{s73}%
\end{align}

By comparing Eq. (\ref{s66}) with (\ref{s70}), it is shown that, when $B$
system passes the laser channel, the upper bound time for the steering of $B$
depends on both loss (gain) factor and squeeing parameter, while that for the
steering of $A$ only depends on loss (gain) factor. In addition, when $B$
system passes only the loss channel, $A$ can always steer $B$ for any long
decoherence time but $B$ can only steer $A$ within a threshold value, i.e.,
$\frac{1}{2}\ln2$. This means that the region of $(0,\frac{1}{2}\ln2)$ is
two-way steering condition, while the range of $(\frac{1}{2}\ln2,\infty)$ is
one-way steering condition.

When $B$ system passes only the gain channel, however, it is interesting that
$B$ can always steer $A$ for any long decoherence time but $A$ can only steer
$B$ within a threshold value, i.e., $\frac{1}{2}\ln \left(
2-\operatorname{sech}^{2}r\right)  $, which is smaller than $\frac{1}{2}\ln2$.
In this sense, the gain channel seems beneficial for the steering of $A,$
while the loss one is for the steering of $B.$ The regions of $(0,\frac{1}%
{2}\ln \left(  2-\operatorname{sech}^{2}r\right)  )$ and $(\frac{1}{2}%
\ln \left(  2-\operatorname{sech}^{2}r\right)  ,\infty)$ are two-way and one
way steering conditions, respectively.

Furthermore, when $B$ system passes the thermal channel, the corresponding
upper bounds of steerable time are given by Eqs. (\ref{s69}) and (\ref{s73})
for the steering of $B$ and $A$, respectively, which increase as $r$ increases
but reduce for higher thermal niose. In particular, it is found that if the
average photon number $\bar{n}$ of thermal noise is equal to half of the total
average photon number $\left \langle n\right \rangle =2\sinh^{2}r$ of the TMSV,
i.e., $\bar{n}=\left \langle n\right \rangle /2$, then both $A$ steering $B$ and
$B$ steering $A$ share the same upper bound of steerable time. If $\bar
{n}>\left \langle n\right \rangle /2$ ($\bar{n}<\left \langle n\right \rangle
/2$), the latter (former) will possess a\ longer steerable time than the
former (latter). This implies that one can modulate the direction of steering
between $A$ and $B$ systems or the two-way steerable condition by changing the
ratio of $\bar{n}$ to $\left \langle n\right \rangle $. This point can be clear
from Fig. 1.

Next, we compare the cases of both one-side and two-side decoherence. (i) For
the case of loss, see Eqs. (\ref{s58}), (\ref{s67}) and (\ref{s71}) it is
found that, a party without loss can steer the other party with loss, while
the case is not true vice verse when $\kappa t$ exceeds a certain threshold
$\frac{1}{2}\ln2$, which is independent of squeezing parameter $r$.\ While
both of two parties pass photon-loss channels, the threshold for two-way
steering is kept unchanged. This indicates that two-side loss has no more
effect on the steerable time than one-side loss. (ii) For the case of gain, it
is shown that, the more \textquotedblleft gained" party can steer the other
one, while the vice verse is not true when $gt$ $>\frac{1}{2}\ln \left(
2-\operatorname{sech}^{2}r\right)  $ depending on $r$, see Eqs. (\ref{s68})
and (\ref{s72}). However, when two parties are \textquotedblleft gained", it
is clear that one-side gain presents a longer steerable time than two-side
gain, see Eqs. (\ref{s60}) and (\ref{s68}). This shows that two-side gain has
more effect on the steerable time than one-side gain. In this sense, the gain
case is in contrast to the loss case. (iii) For the case of thermal noise,
two-side channel presents much shorter steerable time than one-side case, see
Fig. 1.

\bigskip

\section{Qualification of quantum steering in laser channel}

In these discussions above, we have employed two kinds of classical steering
criteria: Reid and entropy criteria. As discussed above, for Gaussian states,
these two criteria are equivalent, which are qualitative steering criteria. In
order to further quantifying the steerability, some measurement methods are
proposed \cite{PhysRevLett.112.180404,PhysRevLett.114.060403,jevtic2015einstein}.
In particular, for two-mode Gaussian states, whose covariance matrix is
characteristic of
\begin{equation}
\Sigma_{AB}=\left(
\begin{array}
[c]{cc}%
\mathcal{A} & \mathcal{C}\\
\mathcal{C}^{T} & \mathcal{B}%
\end{array}
\right)  , \label{s74}%
\end{equation}
where $\mathcal{A},\mathcal{B}$ and $\mathcal{C}$ are $2\times2$ submatrices,
the steeriability of Bob by Alice ($A\rightarrow B$) or Alice by Bob
($B\rightarrow A$) can be qualified by the following quantities%
\begin{align}
\mathcal{G}^{A\rightarrow B}  &  =\max \left \{  0,\frac{1}{2}\ln \frac
{\det \mathcal{A}}{\det \Sigma_{AB}}\right \}  ,\label{s75}\\
\mathcal{G}^{B\rightarrow A}  &  =\max \left \{  0,\frac{1}{2}\ln \frac
{\det \mathcal{B}}{\det \Sigma_{AB}}\right \}  , \label{s76}%
\end{align}
respectively.

Using Eqs. (\ref{s75}) and (\ref{s76}), we can obtain the steeriability for
the cases of both two-side and one-side channels, respectively, as%
\begin{equation}
\mathcal{G}_{2}^{A\leftrightarrow B}=\max \left \{  0,\ln \frac{B}{\allowbreak
B^{2}-C^{2}}\right \}  , \label{s77}%
\end{equation}
and
\begin{align}
\mathcal{G}_{1}^{A\rightarrow B}  &  =\max \left \{  0,\ln \frac{\cosh
2r}{R+A\cosh2r}\right \}  ,\label{s78}\\
\mathcal{G}_{1}^{B\rightarrow A}  &  =\max \left \{  0,\ln \frac{A+R\cosh
2r}{R+A\cosh2r}\right \}  . \label{s79}%
\end{align}
Here, the subscripts $1$ and $2$ represent the cases of one-side and two-side
channels, respectively. Comparing Eqs. (\ref{s77})-(\ref{s79}) with Eqs.
(\ref{s56}), (\ref{s64}) and (\ref{s65}), it is interesting to notice that
they share the same steerable condition, or the same steerable time. In
particular, for photon-gain ($\kappa=0$) and thermal noise ($g\rightarrow
\kappa \bar{n},\kappa \rightarrow \kappa \left(  \bar{n}+1\right)  $) channels,
Eqs. (\ref{s77})-(\ref{s79}) reduce to, respectively,%

\begin{align}
\mathcal{G}_{2}^{A\leftrightarrow B}  &  =\max \left \{  0,\ln \frac{R\cosh
2r-T}{\left(  \allowbreak Re^{2r}-T\right)  \left(  Re^{-2r}-T\right)
}\right \}  ,\label{s80}\\
\mathcal{G}_{1}^{A\rightarrow B}  &  =\max \left \{  0,\ln \frac{1}%
{R\operatorname{sech}2r-T}\right \}  ,\label{s81}\\
\mathcal{G}_{1}^{B\rightarrow A}  &  =\max \left \{  0,\ln \frac
{R-T\operatorname{sech}2r}{R\operatorname{sech}2r-T}\right \}  ,\label{s82}\\
&  \left(  R=e^{2gt},T=1-R\right) \nonumber
\end{align}
and
\begin{align}
\mathcal{G}_{2}^{A\leftrightarrow B}  &  =\max \left \{  0,\ln \frac{NT+R\cosh
2r}{\left(  NT+Re^{2r}\right)  \left(  NT+Re^{-2r}\right)  }\right \}
,\label{s83}\\
\mathcal{G}_{1}^{A\rightarrow B}  &  =\max \left \{  0,\ln \frac{1}%
{R\operatorname{sech}2r+NT}\right \}  ,\label{s84}\\
\mathcal{G}_{1}^{B\rightarrow A}  &  =\max \left \{  0,\ln \frac
{R+NT\operatorname{sech}2r}{NT+R\operatorname{sech}2r}\right \}  .\label{s85}\\
&  \left(  R=e^{-2\kappa t},T=1-R,N=2\bar{n}+1\right) \nonumber
\end{align}
When $N=1$ Eqs. (\ref{s83})-(\ref{s85}) reduce to the case of photon-loss
($g=0$).

\begin{figure}[ptb]
\label{F1} \centering \includegraphics[width=8cm]{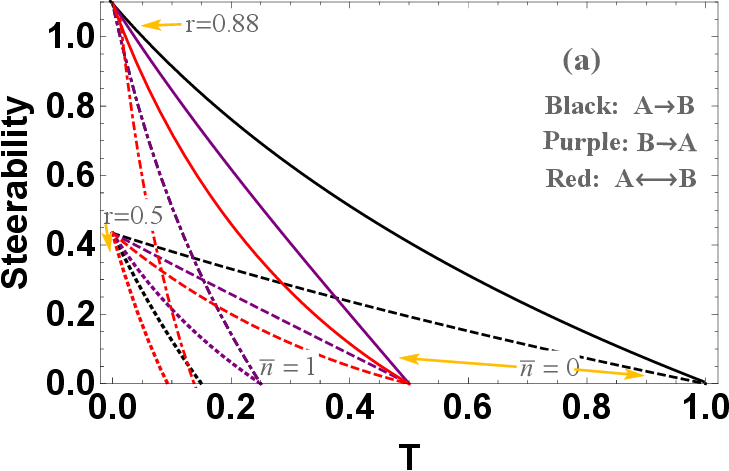} \newline%
\includegraphics[width=8cm]{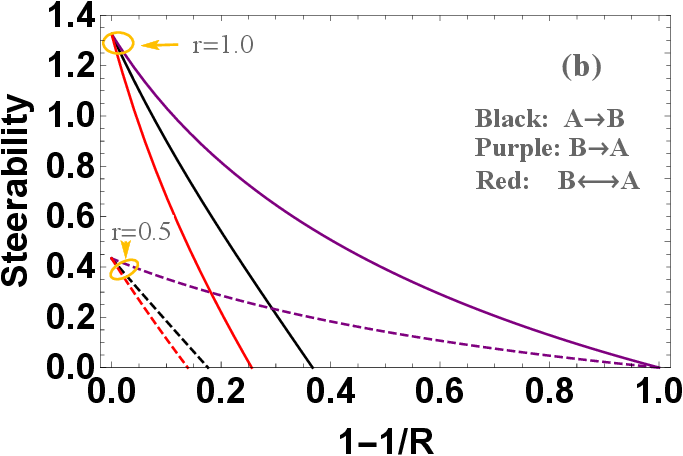}\caption{{}(Color online) The
steerability as a function of $T=1-R$ or $R$ with some given parameters $r$
and $\bar{n},$ (a) for thermal channel including photon-loss channel,
$R=e^{-2\kappa t}$; For same colors, dashed, solid, dotted, dotdashed lines
correspond to $r=0.5,0.88,0.5,0.88,$\ respectively. (b) for gain channel,
$R=e^{2gt}$. Black, purple and red lines correspond to $\mathcal{G}%
_{1}^{A\rightarrow B},$ $\mathcal{G}_{1}^{B\rightarrow A}$ and $\mathcal{G}%
_{2}^{A\leftrightarrow B}$, respectively. }%
\label{Figure2}%
\end{figure}

In order to clearly see the effect of different parameters on the
steerability, we plot the steerability as a function of $T=1-R$ for thermal
channel with some given parameters $r$ and $\bar{n}$ in Fig. 2(a). From Fig.
2(a) it is clear that (i) for a given squeezing parameter $r$ (thermal average
photon $\bar{n}$)$,$ the steerability of $\mathcal{G}_{2}^{A\leftrightarrow
B}$ (black), $\mathcal{G}_{1}^{A\rightarrow B}$ (purple) and $\mathcal{G}%
_{1}^{B\rightarrow A}$ (red) decrease (increase) with the increase of $\bar
{n}$ ($r$); (ii) for two-side and one-side photon-loss cases, although they
share the same two-way steerable time (say see solid purple and solid red
lines), their steerabilities are different from each other, i.e., two-side
loss presents a smaller steerability than one-side loss, as expected. This
implies that two-side loss may destroy the system coherence more seriously
than one-side loss.

In Fig. 2(b), for gain channel we plot steerability as a function of
$1-1/R=1-e^{-2gt}$. It is shown that (i) the more gained party can steer the
other's state, while the other party cannot steer in a certain threshold
value, which depends on the squeezing parameter $r$. This case is in sharp
contrast to that the less decohered party can steer the other's state, while
the other party cannot steer. In this sense, it seems that the gain effect in
one party is equivalent to the loss effect in the other's party. (ii) In terms
of steerable length, it can be arranged in the following chain inequalities
for given parameter $r$, $\mathcal{G}_{1}^{B\rightarrow A}>\mathcal{G}%
_{1}^{A\rightarrow B}>\mathcal{G}_{2}^{A\leftrightarrow B}$. This implies that
two-side gain channel presents much worse steerability due to the fact that
the two-side gain may further destroy the coherence.

In general, for the laser channel, $\kappa \neq g\neq0$, thus for simplicity we
take $g/\kappa=\gamma$. In order to see clearly the effect of $g/\kappa$ on
the steerability, for given $r$ and different $\gamma$ parameters we plot the
steerability as a function of dimensionless time $\kappa t\ $in Fig. 3. It is
found that (i) there is always a certain threshold time $\kappa t_{c}$ when
both gain and loss factors are present for all of $\mathcal{G}_{1}%
^{B\rightarrow A},\mathcal{G}_{1}^{A\rightarrow B}\ $and $\mathcal{G}%
_{2}^{A\leftrightarrow B},$ even for the case of $\kappa=g$. (ii) For a given
$\gamma$, the threshold can be arranged in the following chain inequalities,
i.e., $t_{c}^{A\rightarrow B}>t_{c}^{B\rightarrow A}>t_{c}^{A\leftrightarrow
B}$. This indicates that there is always a threshold time for $\mathcal{G}%
_{1}^{B\rightarrow A}$ and $\mathcal{G}_{1}^{A\rightarrow B}$ when both gain
and loss are present in one side, which is different from the cases only with
loss or gain. Gain and loss in two-side further shorten the threshold. (iii)
In addition, for $\mathcal{G}_{1}^{B\rightarrow A}\ $or $\mathcal{G}%
_{1}^{A\rightarrow B}\ $or $\mathcal{G}_{2}^{A\leftrightarrow B}$, the
corresponding threshold $\kappa t_{c}$ decreases as the increase of $\gamma$.
This implies that a larger gain $g$ will lead to a shorter threshold time.

\begin{figure}[ptb]
\label{F3} \centering \includegraphics[width=8cm]{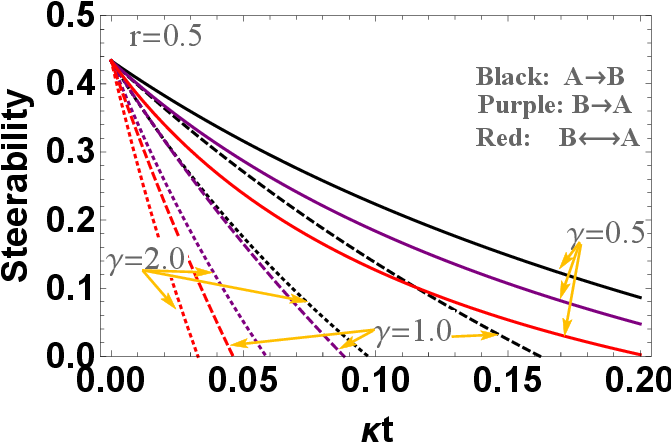} \newline%
\caption{{}(Color online) The steerability as a function of $\kappa t$ with
some given parameters $r=0.5$ and $g/\kappa=\gamma=0.5,1.0.2.0.$\ Black,
purple and red lines correspond to $\mathcal{G}_{1}^{A\rightarrow B},$
$\mathcal{G}_{1}^{B\rightarrow A}$ and $\mathcal{G}_{2}^{A\leftrightarrow B}$,
respectively. For same colors, solid, dashed, dotted, lines correspond to
$\gamma=0.5,1.0.2.0,$\ respectively.}%
\end{figure}

\section{Comparing with the decoherence of quantum entanglement}

Quantum steering is a kind of quantum correlation, which is closely related to
quantum entanglement and Bell nonlocality. For Gaussian states, the relation
between quantum steering and entanglement has been discussed in Refs.\cite{he2015classifying}. In this section, we further make a
comparison about the decoherence between quantum steering and entanglement in
laser channel.

For two-mode Gaussian states, the degree of entanglement can be calculated by
the logarithmic negativity ,
\begin{equation}
E_{N}=\max[0,-\ln \nu_{s}], \label{s86}%
\end{equation}
where $\nu_{s}$ is the smallest symplectic eigenvalue of partially transposed
state. Here the symplectic eigenvalues can be defined as the modulus of the
eigenvalues of a matrix $i\Omega \Sigma$, where $\Sigma$ is the covariance
matrix (CM) of partially transposed state and $\Omega=\left(
\begin{array}
[c]{cc}%
0 & 1\\
-1 & 0
\end{array}
\right)  \oplus \left(
\begin{array}
[c]{cc}%
0 & 1\\
-1 & 0
\end{array}
\right)  $.

Next, we compare the decoherence of entanglement and steering under two-side
and one-side channels. First, we consider the two-side case. For this purpose,
from Eq. (\ref{s47}) the CM is given by%
\begin{equation}
\Sigma_{2}=\left(
\begin{array}
[c]{cccc}%
B & 0 & C & 0\\
0 & B & 0 & -C\\
C & 0 & B & 0\\
0 & -C & 0 & B
\end{array}
\right)  . \label{s87}%
\end{equation}
The symplectic eigenvalues can be calculated as $\nu_{1}=B+C,\nu_{2}=B-C,$
then $\nu_{s}=\min[\nu_{1},\nu_{2}]=B-C$, such that $E_{N}=\max[0,-\ln(B-C)]$.
Thus the unseperable condition is given by $B-C<1.$ On the other hand, under
the condition of two-side channels, from Eq. (\ref{s55}) the steerable
condition is given by $B-C<B/(B+C)<1$. Comparing these two conditions, it is
interesting to notice that the inseperable condition must be satisfied when
the steerable condition is satisfied, while the case is not true vice verse.
This implies that the steerable condition is stronger\ than the unseperable
condition for describing quantum correlation, or the steerable time is shorter
than the unseperable time. In fact, it is ready to have a critical time for
the inseperablity
\begin{equation}
t<t_{c}=\frac{1}{2\left(  \kappa-g\right)  }\ln \frac{g+\kappa \tanh r}{g\left(
1+\tanh r\right)  }, \label{s88}%
\end{equation}
as expected \cite{PhysRevA.76.012333}. In particular, for photon-loss, photon-gain and
thermal channels, Eq. (\ref{s88}) reduces to, respectively,
\begin{align}
\kappa t_{c}  &  \rightarrow \infty,\label{s89}\\
gt_{c}  &  =\frac{1}{2}\ln \left(  1+\tanh r\right)  ,\label{s90}\\
\kappa t_{c}  &  =\frac{1}{2}\ln \left(  1+\frac{1/\bar{n}\tanh r}{1+\tanh
r}\right)  , \label{s91}%
\end{align}
Eq. (\ref{s91}) is also obtained in Ref. \cite{PhysRevLett.84.2722}. These critical
thresholds in Eqs. (\ref{s89})-(\ref{s91}) are longer than those for
steerability, as expected.

Now, we further examine the single-side case. In a similar way, the
corresponding degree of entanglement is given by%
\begin{equation}
E_{N}=\max \left \{  0,-\ln \left[  B_{+}-\left(  B_{-}^{2}+\frac{C^{2}}%
{R}\right)  ^{1/2}\right]  \right \}  , \label{s92}%
\end{equation}
where $B_{\pm}=(B\pm \cosh2r)/2$. At initial time $t=0$, Eq. (\ref{s92})
becomes $\ln e^{2r}$, as expected. Thus the inseperable condition can be
derived as%
\begin{equation}
t<t_{c}=\frac{1}{2\left(  \kappa-g\right)  }\ln \frac{\kappa}{g}. \label{s93}%
\end{equation}
It is interesting to notice that the inseperable condition is independent of
squeezing parameter $r>0$ when $B$-mode passes the laser channel. This case is
similar to the case of $A\rightarrow B$ steering but different from that of
$B\rightarrow A$ steering. In addition, the inseperable threshold decreases as
the increase of gain factor $g$. In particular, for photon-loss, photon-gain
and thermal channels, Eq. (\ref{s93}) reduces to $\kappa t_{c}\rightarrow
\infty,gt_{c}\rightarrow \infty,$ and $\kappa t_{c}=\frac{1}{2}\ln \frac{\bar
{n}+2}{\bar{n}}$, respectively. Compared with Eqs. (\ref{s67})-(\ref{s69}) and
(\ref{s71})-(\ref{s73}), it is ready to see that the steerable condition can
be seen as a subset of the inseperable condition, as expected. Actually, this
case can be generally checked by comparing Eq. (\ref{s93}) and Eqs.
(\ref{s66}) and (\ref{s70}).

 \begin{figure}[ptb]
\label{F4} \centering \includegraphics[width=8cm]{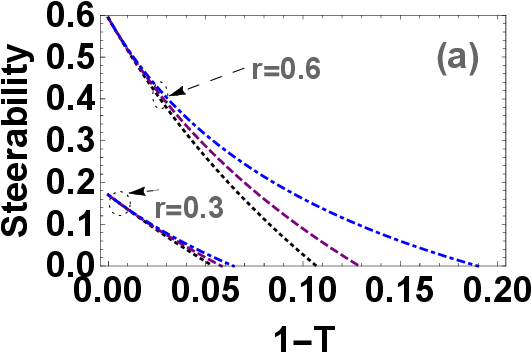} \newline%
\includegraphics[width=8cm]{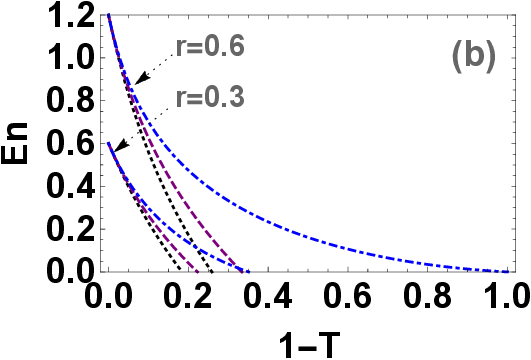}\caption{{}(Color online) The
steerability and entanglement as a function of $1-T$ with some given
parameters $r=0.3,0.6,$ $\bar{n}=1$ and $M=0,1,\sqrt{2}$ (corresponding to
black dotted, purple dashed, and blue dotdashed, respectively), (a) the degree
of entanglement, (b) the steerability in two-side channel. }%
\end{figure}

\begin{figure}[ptb]
\label{F5} \centering \includegraphics[width=8cm]{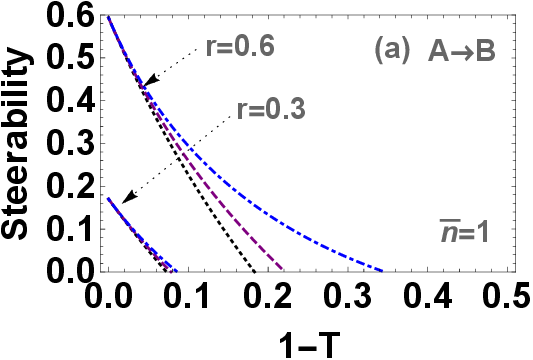} \newline%
\includegraphics[width=8cm]{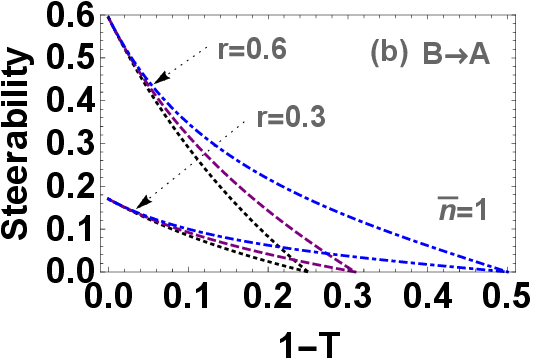}\newline%
\includegraphics[width=8cm]{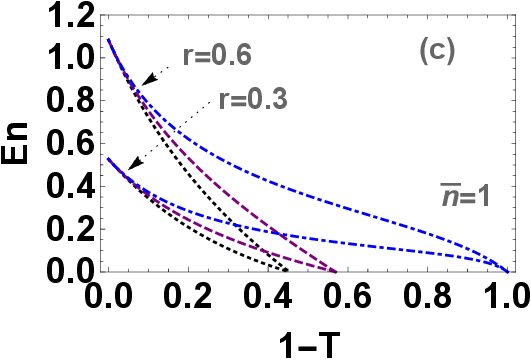}\caption{{}(Color online) The
steerability and entanglement as a function of $1-T$ with some given
parameters $r=0.3,0.6,$ $\bar{n}=1$ and $M=0,1,\sqrt{2}$ (corresponding to
black dotted, purple dashed, and blue dotdashed, respectively), (a)-(b) the
steerability $A\rightarrow B$ and $B\rightarrow A$; (b) the degree of
entanglement. }%
\end{figure}
\section{Discussions under phase-sensitive enviornment }

Actually, our dicsussions can be directly extended to the general noise
channel, i.e., next we consider the evolution of the TMSV in the presence of
phase-sensitive enviornments, which can be described as the following master
equation for single-mode case \cite{PhysRevA.69.022318}%

\begin{align}
\frac{d\rho}{dt}  &  =\kappa \left(  \bar{n}+1\right)  \left(  2a\rho a^{\dag
}-a^{\dag}a\rho-\rho a^{\dag}a\right) \nonumber \\
&  +\kappa \bar{n}\left(  2a^{\dag}\rho a-aa^{\dag}\rho-\rho aa^{\dag}\right)
\nonumber \\
&  -\kappa M\left(  2a^{\dag}\rho a^{\dag}-a^{\dag2}\rho-\rho a^{\dag2}\right)
\nonumber \\
&  -\kappa M^{\ast}\left(  2a\rho a-a^{2}\rho-\rho a^{2}\right)  , \label{s94}%
\end{align}
where $M$ is a complex parameter related to the squeezed enviornments,
satisfying $\left \vert M\right \vert ^{2}\leqslant \bar{n}\left(  \bar
{n}+1\right)  $ for a qualified density operator. In a similar way to Eq.
(\ref{s44}), using the entangled state representation, one can obtain the
evolution of the CF \cite{chen2012new} (see Appendix B) as%
\begin{equation}
\chi \left(  \alpha,t\right)  =e^{-\frac{R}{2}(N\left \vert \alpha \right \vert
^{2}-M\alpha^{\ast2}-M^{\ast}\alpha^{2})}\chi(\alpha \sqrt{T},0) \label{s95}%
\end{equation}
with $N=2\bar{n}+1,R=1-e^{-2\kappa t},T=e^{-2\kappa t}$, and $\chi \left(
\alpha,0\right)  $ is the initial CF. In particular, for the initial Gaussian
state, the CF can be given by \cite{RevModPhys.84.621}%
\begin{equation}
\chi(\alpha,0)=\exp \left \{  -\frac{1}{2}\xi \Omega V_{0}\Omega^{T}\xi
^{T}\right \}  , \label{s96}%
\end{equation}
where $\xi=$ $\left(
\begin{array}
[c]{cc}%
q & p
\end{array}
\right)  $ and $V_{0}$ is the covariance matrix of the initial Gaussian state,
as well as $\Omega=\left(
\begin{array}
[c]{cc}%
0 & 1\\
-1 & 0
\end{array}
\right)  .$ Here for simplicity, we have assumed $\bar{\xi}=0$. In this case,
after substituting Eq. (\ref{s96}) into Eq. (\ref{s95}), the evolution of the
covariance matrix is shown as%
\begin{equation}
V_{out}=RV_{\infty}+TV_{0}, \label{s97}%
\end{equation}
where $V_{\infty}=\left(
\begin{array}
[c]{cc}%
N+2\operatorname{Re}M & 2\operatorname{Im}M\\
2\operatorname{Im}M & N-2\operatorname{Re}M
\end{array}
\right)  $ \ is the covariance matrix of the output state when $t\rightarrow
\infty$ \cite{PhysRevA.57.548,PhysRevA.68.012314,PhysRevA.52.2401}.

Next, we consider the cases that the TSMV pass through single-side and
two-side phase-sensitive channels, reespctively. Using Eq. (\ref{s97}), we can
get the covariance matrix of the output state for these two cases above, which
has the following form%

\begin{equation}
V_{j}=\left(
\begin{array}
[c]{cc}%
\mathcal{A}_{j} & \mathcal{C}_{j}\\
\mathcal{C}_{j}^{T} & \mathcal{B}_{j}%
\end{array}
\right)  , \label{s98}%
\end{equation}
where, the subscripts $j=1$, $2$ represent the cases of single-side and
two-side channels, respectively. The entanglement degree described by Eq.
(\ref{s98}) is given by Eq. (\ref{s86}), where $2\nu_{s}^{2}=\Delta
-\sqrt{\Delta^{2}-4\det V_{j}}$, and $\Delta=\det \mathcal{A}_{j}%
+\det \mathcal{B}_{j}-2\det \mathcal{C}_{j}$.

Specifically speaking, for single- and two-side decoherence, the CMs are,
respectively, given by%
\begin{equation}
V_{1}=\left(
\begin{array}
[c]{cccc}%
a_{1} & 0 & d_{1} & 0\\
0 & a_{1} & 0 & -d_{1}\\
d_{1} & 0 & b_{1}+c_{1} & f_{1}\\
0 & -d_{1} & f_{1} & b_{1}-c_{1}%
\end{array}
\right)  , \label{s99}%
\end{equation}
with $a_{1}=\cosh2r,b_{1}=T\cosh2r+RN,c_{1}=2R\operatorname{Re}M,d_{1}%
=\sqrt{T}\sinh2r,f_{1}=2R\operatorname{Im}M$, and
\begin{equation}
V_{2}=\left(
\begin{array}
[c]{cccc}%
a_{2} & c_{2} & d_{2} & 0\\
c_{2} & b_{2} & 0 & -d_{2}\\
d_{2} & 0 & a_{2} & c_{2}\\
0 & -d_{2} & c_{2} & b_{2}%
\end{array}
\right)  , \label{s100}%
\end{equation}
with $a_{2}=R\left(  N+2\operatorname{Re}M\right)  +T\cosh2r$, $b_{2}=R\left(
N-2\operatorname{Re}M\right)  +T\cosh2r$, $c_{2}=2R\operatorname{Im}M$,
$d_{2}=T\sinh2r$. Thus the steerability can be calculated as
\begin{align}
\mathcal{G}_{1}^{A\rightarrow B}  &  =\max \left \{  0,\frac{1}{2}\ln \frac
{a_{1}^{2}}{\allowbreak \det V_{1}}\right \}  ,\label{s101}\\
\mathcal{G}_{1}^{B\rightarrow A}  &  =\max \left \{  0,\frac{1}{2}\ln
\frac{\allowbreak b_{1}^{2}-c_{1}^{2}-f_{1}^{2}}{\allowbreak \det V_{1}%
}\right \}  ,\label{s102}\\
\mathcal{G}_{2}^{A\leftrightarrow B}  &  =\max \left \{  0,\frac{1}{2}\ln
\frac{a_{2}b_{2}-c_{2}^{2}}{\det V_{2}}\right \}  . \label{s103}%
\end{align}
In particular, when $\bar{n}=0$ ($M=0$) and $M=0,$ both entanglement and
steerability do reduce to the previous reslults.

In order to examine the effects of phase-sensitive enviornments, especially
the squeezing parameter $M$ on both entanglement and steerability, we plot the
entanglement and steerability as a function of $1-T$ for several given
parameters, $r$, $\bar{n}$ and $M$ in Figs. 4 and 5. Figs. 4 and 5 present the
cases of two-side symmetrical decoherence channel, and single side decoherence
channel, respectively. For both of two cases, the existence of thermal noise
will further reduce both entanglement and steerability, and the reduced effect
will becomes obvious with the increase of $\bar{n}$. For a given $\bar{n}=1$,
the case with a bigger squeezing parameter $r$ will be benificial to enhancing
both entanglement and steerability. However, it presents a faster reduction
due to the decoherence when compared to the case with a smaller squeezing
parameter. This case is true when the squeezing enviornments are involved.
Under this condition of $\left \vert M\right \vert ^{2}\leqslant \bar{n}(\bar
{n}+1)$, in addition, it is interesting to notice that the larger the
parameter $M$ is, the more obvious the enhanced effects of the entanglement
and steerability is. Here for simplicity, we take $M$ real. At a fixed $M,$
the enhancement are more effective for the case with a larger squeezing
parameter $r$. These above are true for both single-side and two-side
decoherence cases.

Comparing Figs. 4 with Fig. 5, it is clear that both entanglement and
steerability present the worst performance under two-side decoherence case.
Although the the parameter $M$ can be used to enhance the entanglement and
steerability for both cases, the enhanced effect is more obvious for
single-side case (see Fig. 5(a)-(c)). In particular, from Fig. 5(c), it is
clear that a same decoerence time is almost shared by different squeezing
parameters ($r=0.3,0.6$) when $M=\sqrt{2}$($\bar{n}=1$), although the case
with $r=0.3$ presents a smaller degree of entanglement than that with $r=0.6$.
It is interesting to notice that the decoerence time is the same as that
without thermal niose, which implies that the introduction of parameter $M$
seems to undo the decoherence effect from the thermal noise. This case is also
true for the steering of $A$ rather than $B$ (see Fig. 5(a) and (b)). These
results mean that when only $B$ passes through the channel, (i) the two-way
steerable time can be prolonged, which increases with the parameter $M;$ (ii)
even for the steering of $A$, the effect of thermal noise seems to be
completely eliminated in the limit of $\left \vert M\right \vert ^{2}=\bar
{n}(\bar{n}+1)$, which is not true for the steering of $B.$

\section{Conclusions}
 In summary, we have  studied the quantum EPR steering for two-mode states with
continuous-variable in laser channel by using the representation of characteristic function. By setting different channel parameters, we can simultaneously analyse the evolution of EPR Steering in loss, gain, and thermal channels. Concretely, it shows that the bound of two-way steering time not only depends on the gain
and loss factors, but also on the squeezing parameter in laser channel, while the squeezing parameter does not influence the bound in the case of loss channel. Furthermore, we found the gain process will introduce additional noise to the two-mode squeezed vacuum states such that the steerable time is reduced when they under one-mode and two-mode laser channel respectively. When $B$ system passes only the gain channel, we found that $B$ can always steer $A$ for any long decoherence time but $A$ can only steer $B$ within a threshold value, which is opposite with loss channel. In addition, by comparing the steerable and inseparable conditions, it is shows that the steerable condition can be seen as a subset of the inseparable condition. At last, we extended our discussions to the phase-sensitive
environment. Our results are of great significance for the design of effective EPR steering distillation schemes in practical quantum channels.

This work is supported by the National Natural Science
Foundation of China (Grants No. 11964013, 61975077 ) and the National Key Research and Development Program of China (Grant Nos.2019YFA0308704, 2018YFA0306202), the Training Program for Academic and Technical Leaders of Major Disciplines in Jiangxi Province (No. 20204BCJL22053).

\bigskip

\bigskip

\bigskip

\bigskip

\textbf{Appendix A: Deriviation of Eq. (\ref{s44})}

Here we derive Eq. (\ref{s44}) by using the thermal entangled state
representation method. For this purpose, we intorduce a state vector, denoted
as $\left \vert I\right \rangle =\exp \left \{  a^{\dag}\tilde{a}^{\dag}\right \}
\left \vert 0\tilde{0}\right \rangle =\sum_{m}^{\infty}\left \vert m,\tilde
{m}\right \rangle $, where $m=\tilde{m},$ $a^{\dag}$ and $\tilde{a}^{\dag}$ are
the creation operators of real mode and fictitious mode, respectively,
satisfying $[\tilde{a},\tilde{a}^{\dag}]=1$ and $\tilde{a}\left \vert
0\tilde{0}\right \rangle =0$, then the thermal entangled state representation
can be obtained by the displacement for real mode, i.e.,
\begin{align}
\left \vert \eta \right \rangle  &  =D_{a}\left(  \eta \right)  \left \vert
0\tilde{0}\right \rangle \nonumber \\
&  =e^{-\left \vert \eta \right \vert ^{2}/2+\eta a^{\dag}-\eta^{\ast}\tilde
{a}^{\dag}+a^{\dag}\tilde{a}^{\dag}}\left \vert 0\tilde{0}\right \rangle ,
\tag{A1}%
\end{align}
where $D_{a}\left(  \eta \right)  =\exp \left \{  \eta a^{\dag}-\eta^{\ast
}a\right \}  $ is the displacement operator for mode $a.$\ It is not difficult
to get the following eigen-equations:
\begin{align}
\left(  a-\tilde{a}^{\dag}\right)  \left \vert \eta \right \rangle  &
=\eta \left \vert \eta \right \rangle ,\left(  a^{\dag}-\tilde{a}\right)
\left \vert \eta \right \rangle =\eta^{\ast}\left \vert \eta \right \rangle
,\tag{A2}\\
\left \langle \eta \right \vert \left(  a^{\dag}-\tilde{a}\right)   &
=\eta^{\ast}\left \langle \eta \right \vert ,\left \langle \eta \right \vert \left(
a-\tilde{a}^{\dag}\right)  =\eta \left \langle \eta \right \vert , \tag{A3}%
\end{align}
with $\left[  \left(  a-\tilde{a}^{\dag}\right)  ,\left(  a^{\dag}-\tilde
{a}\right)  \right]  =0$. When $\eta=0,$ from Eqs. (A2) and (A3) it is ready
to have%
\begin{equation}
a\left \vert I\right \rangle =\tilde{a}^{\dag}\left \vert I\right \rangle
,\tilde{a}\left \vert I\right \rangle =a^{\dag}\left \vert I\right \rangle
,a^{\dag}a\left \vert I\right \rangle =\tilde{a}^{\dag}\tilde{a}\left \vert
I\right \rangle . \tag{A4}%
\end{equation}

Next, we bridge the relation between the characteristic function (CF) and
$\left \langle \eta \right \vert $. According to the CF definition of density
operator $\rho$, i.e., $\chi \left(  \lambda \right)  =\mathtt{tr}(e^{\lambda
a^{\dag}-\lambda^{\ast}a}\rho),$ then the CF can be reformed to the following
form%
\begin{align}
\chi \left(  \lambda \right)   &  =\sum_{m,n}^{\infty}\left \langle n,\tilde
{n}\right \vert e^{\lambda a^{\dag}-\lambda^{\ast}a}\rho \left \vert m,\tilde
{m}\right \rangle \nonumber \\
&  =\left \langle I\right \vert D_{a}\left(  \lambda \right)  \left \vert
\rho \right \rangle \nonumber \\
&  =\left \langle \eta_{=-\lambda}\right \vert \left.  \rho \right \rangle
,\tag{A5}%
\end{align}
where $\left \vert \rho \right \rangle \equiv \rho \left \vert I\right \rangle ,$
$m=\tilde{m},$ and $n=\tilde{n}$.

In order to derive Eq. (\ref{s44}), operating Eq. (\ref{s43}) on $\left \vert
I\right \rangle $ and using Eq. (A4), we have %
\begin{align}
\frac{d}{dt}\left \vert \rho \left(  t\right)  \right \rangle  &  =[g\left(
2a^{\dag}\tilde{a}^{\dag}-aa^{\dag}-\tilde{a}\tilde{a}^{\dag}\right)
\nonumber \\
&  +\kappa \left(  2a\tilde{a}-a^{\dag}a-\tilde{a}^{\dag}\tilde{a}\right)
]\left \vert \rho \left(  t\right)  \right \rangle \nonumber \\
&  =[\left(  \kappa+g\right)  \left(  \tilde{a}-a^{\dag}\right)  \left(
a-\tilde{a}^{\dag}\right) \nonumber \\
&  +\left(  \kappa-g\right)  \left(  a\tilde{a}-a^{\dag}\tilde{a}^{\dag
}+1\right)  ]\left \vert \rho \left(  t\right)  \right \rangle . \tag{A6}%
\end{align}
After operating Eq. (A6) from left-side on $\left \langle \eta_{=-\lambda
}\right \vert $, and using Eqs. (A2)-(A3) as well as
\begin{align}
\left \langle \eta \right \vert \tilde{a}  &  =-\left(  \frac{\partial}%
{\partial \eta}+\frac{\eta^{\ast}}{2}\right)  \left \langle \eta \right \vert
,\tag{A7}\\
\left \langle \eta \right \vert \tilde{a}^{\dag}  &  =\left(  \frac{\partial
}{\partial \eta^{\ast}}-\frac{\eta}{2}\right)  \left \langle \eta \right \vert
,\tag{A8}\\
\left \langle \eta \right \vert a  &  =\left(  \frac{\partial}{\partial \eta
^{\ast}}+\frac{\eta}{2}\right)  \left \langle \eta \right \vert ,\tag{A9}\\
\left \langle \eta \right \vert a^{\dag}  &  =-\left(  \frac{\partial}%
{\partial \eta}-\frac{\eta^{\ast}}{2}\right)  \left \langle \eta \right \vert ,
\tag{A10}%
\end{align}
it is ready to have%
\begin{align}
\frac{d}{dt}\chi \left(  \lambda \right)   &  =[-\left(  \kappa-g\right)
(\eta \frac{\partial}{\partial \eta}+\eta^{\ast}\frac{\partial}{\partial
\eta^{\ast}})\nonumber \\
&  -\left(  \kappa+g\right)  \left \vert \eta \right \vert ^{2}]\left \langle
\eta \right.  \left \vert \rho \left(  t\right)  \right \rangle , \tag{A11}%
\end{align}
whose solution is given by%
\begin{equation}
\chi \left(  \lambda,t\right)  =e^{-\left(  \kappa+g\right)  t\left \vert
\eta \right \vert ^{2}-\left(  \kappa-g\right)  t(\eta \frac{\partial}%
{\partial \eta}+\eta^{\ast}\frac{\partial}{\partial \eta^{\ast}})}\left \langle
\eta \right.  \left \vert \rho \left(  0\right)  \right \rangle , \tag{A12}%
\end{equation}
where $\eta=-\lambda,$ and $\rho \left(  0\right)  $ is the density operator at
initial time.

Noting $[\eta \frac{\partial}{\partial \eta}+\eta^{\ast}\frac{\partial}%
{\partial \eta^{\ast}},\eta^{\ast}\eta]=2\eta \eta^{\ast},$ and using the
following operator identity
\begin{equation}
e^{f\left[  A+\sigma B\right]  }=e^{\sigma B\left(  e^{f\tau}-1\right)  /\tau
}e^{fA}, \tag{A13}%
\end{equation}
when $\left[  A,B\right]  =\tau B$ is hold, Eq. (A12) can be put into the form%
\begin{align}
\chi \left(  \lambda,t\right)   &  =e^{-\frac{1}{2}\frac{\kappa+g}{\kappa
-g}(1-e^{-2\left(  \kappa-g\right)  t})\left \vert \eta \right \vert ^{2}%
}\nonumber \\
&  \times e^{-\left(  \kappa-g\right)  t\left(  \eta \frac{\partial}%
{\partial \eta}+\eta^{\ast}\frac{\partial}{\partial \eta^{\ast}}\right)
}\left \langle \eta \right.  \left \vert \rho \left(  0\right)  \right \rangle
\nonumber \\
&  =e^{-\frac{1}{2}\frac{\kappa+g}{\kappa-g}(1-e^{-2\left(  \kappa-g\right)
t})\left \vert \eta \right \vert ^{2}}\chi(\lambda e^{-\left(  \kappa-g\right)
t},0), \tag{A14}%
\end{align}
where $\left \langle \eta \right.  \left \vert \rho \left(  0\right)
\right \rangle =\chi \left(  \lambda,0\right)  $ is the initial CF, and we have
used
\begin{align}
&  \exp \left \{  -f\left(  \eta \frac{\partial}{\partial \eta}+\eta^{\ast}%
\frac{\partial}{\partial \eta^{\ast}}\right)  \right \}  \left \langle
\eta \right \vert \nonumber \\
&  =e^{f}\left \langle \eta \right \vert \exp \left \{  f\left(  a\tilde{a}%
-a^{\dag}\tilde{a}^{\dag}\right)  \right \} \nonumber \\
&  =\left \langle \eta e^{-f}\right \vert . \tag{A15}%
\end{align}
Eq. (A14) is just Eq. (\ref{s44}), as expected.

\textbf{Appendix B: Deriviation of Eq. (\ref{s95})}

In a similar way to deriving Eq. (\ref{s44}), we first convert Eq. (\ref{s94})
to
\begin{align}
\frac{d\left \vert \rho \right \rangle }{dt}  &  =\kappa \{ \left(  \bar
{n}+1\right)  \left[  \left(  a-\tilde{a}^{\dag}\right)  \tilde{a}+\left(
\tilde{a}-a^{\dag}\right)  a\right] \nonumber \\
&  +\bar{n}\left[  \left(  a^{\dag}-\tilde{a}\right)  \tilde{a}^{\dag}+\left(
\tilde{a}^{\dag}-a\right)  a^{\dag}\right] \nonumber \\
&  -M\left(  a^{\dag}-\tilde{a}\right)  ^{2}-M^{\ast}\left(  \tilde{a}^{\dag
}-a\right)  ^{2}\} \left \vert \rho \right \rangle , \tag{A16}%
\end{align}
then using Eqs. (A7)-(A10) the CF equation is given by%
\begin{equation}
\frac{d}{dt}\chi \left(  \lambda,t\right)  =-\kappa \lbrack G+\eta \frac
{\partial}{\partial \eta}+\eta^{\ast}\frac{\partial}{\partial \eta^{\ast}}%
]\chi \left(  \lambda,t\right)  , \tag{A17}%
\end{equation}
where $G=\left(  2\bar{n}+1\right)  \left \vert \eta \right \vert ^{2}%
-M\eta^{\ast2}-M^{\ast}\eta^{2}$. Using Eqs. (A13) and (A15) again, we have
\begin{equation}
\chi \left(  \lambda,t\right)  =\exp \left[  -\frac{1}{2}(1-e^{-2\kappa
t})G\right]  \chi \left(  \lambda e^{-\kappa t},0\right)  , \tag{A18}%
\end{equation}
as expected.

$\allowbreak$

\end{document}